\documentstyle[12pt,epsf]{article}
\begin{document}
\title{The Wilson Loop in Yang-Mills Theory  \\
       in the General Axial Gauge}
\author{\large Brian J.
Hand\thanks{bhand@sgi1.mathstat.uoguelph.ca}\, \\
        Dept. of Mathematics and Statistics, University of Guelph
\\ Guelph, Ontario, Canada N1G 2W1 \\
      George Leibbrandt\thanks{gleibbra@msnet.mathstat.uoguelph.ca}
\thanks{Permanent address: Dept. of Math \& Stats.,
U. of Guelph, Guelph, ON N1G 2W1} \\
      Theoretical Physics Division, C.E.R.N., CH-1211 \\
Geneva, Switzerland}
\maketitle
\vspace{-4in}
\begin{flushright}
CERN-TH. 227-95
\end{flushright}
\vspace{3.4in}
\begin{abstract}
We test the unified-gauge formalism by computing a Wilson loop in
Yang-Mills theory to one-loop order.  The unified-gauge formalism
is characterized by the abritrary, but fixed, four-vector $N_\mu$,
which collectively represents the light-cone gauge $(N^2 = 0)$, the
temporal gauge $(N^2 > 0)$, the pure axial gauge $(N^2 < 0)$ and
the planar gauge $(N^2 < 0)$.  A novel feature of the calculation
is the use of distinct sets of vectors, $\{ n_{\mu}, n_{\mu}^{\ast}
\}$ and $\{N_{\mu}, N_{\mu}^{\ast}\}$, for the path and for the
gauge-fixing constraint, respectively.  The answer for the Wilson
loop is independent of $N_{\mu}$, and agrees numerically with the
result obtained in the Feymman gauge.
\end{abstract}

\section{Introduction}

The Wilson loop has proven to be an excellent framework for testing
the consistency of axial-type gauges.  In 1982, Caracciolo, Curci
and Menotti [1] computed the Wilson loop to demonstrate that the
principal-value prescription fails for the temporal gauge, $A_0 =
0$, in both Abelian and non-Abelian gauge theories.  It was later
shown [2,3] in the context of the unified-gauge formalism, that the
$n_{\mu}^{\ast}$-prescription [4,5] does give the correct result
for
the Wilson loop.  The unified-gauge formalism was developed several
years ago by one of the authors [6,7], and tested in detail for the
two-loop Yang-Mills self-energy [8].

In 1989, H\H{u}ffel, Landshoff and Taylor carried out a successful
test of the unified-gauge prescription by demonstrating that the
time dependence of a typical Wilson loop exponentiates to order
$g^4$ [9].  The path in Figure 1 has been used in several previous
computations of the Wilson loop.  For instance, Korchemskaya and
Korchemsky [10], employing dimensional regularization, examined the
Wilson loop to second order perturbation theory in the Feynman
gauge.  In 1992, Andr\u{a}si and Taylor [2] evaluated the same
Wilson loop in the light-cone gauge, suggesting a breakdown of the
$n_{\mu}^{\ast}$-prescription.  However, a detailed analysis by
Bassetto and his co-workers subsequently revealed the absence of
any inconsistencies in the $n_{\mu}^{\ast}$-prescription [3].  In
fact, their light-cone gauge result for the Wilson loop turned out
to be in complete agreement with the corresponding calculation in
the Feynman gauge.

In axial-type gauges, the Lagrangian density for
massless Yang-Mills theory is given by (notice that $n_{\mu}$ in
the preceding paragraphs is now replaced by the letter $N_{\mu}$)

\begin{equation}
L_{YM} = -\frac{1}{4} (F_{\mu\nu}^a)^2 -\frac{1}{2\alpha}
(N\cdot A^a)^2, \alpha\rightarrow 0,
\label{eq:ldYM}
\end{equation}
where $N_{\mu} = (N_0, \bf{N})$ is the gauge-fixing vector, and
\begin{equation}
N^{\mu} A_{\mu}^{a} = 0, \mu = 0,1,2,3,
\end{equation}
the gauge-fixing constraint.

The gauge-field propagator, with gauge indices omitted, reads

\begin{equation}
G_{\mu\nu}(q) = \frac{-i}{q^2 + i\epsilon} \left[ g_{\mu\nu} -
\frac{(q_{\mu}N{\nu} + q_{\nu}N_{\mu})}{q\cdot N} + (N^2 + \alpha
q^2) \frac{q_{\mu}q_{\nu}}{(q\cdot N)^2} \right],
\label{eq:Nprop}
\end{equation}
where $\epsilon > 0$ and $\alpha\rightarrow 0$.

We shall treat the poles of $(q\cdot N)^{-1}$ and $(q\cdot N)^{-2}$
in Eq. (3) with the unified-gauge prescription [6,7], which is a
generalization of the light-cone gauge prescription developed by
Mandelstam [4] and one of the authors [5]:
\begin{eqnarray}
\frac{1}{q\cdot N} \left|^{l.c.} = \lim_{\epsilon\rightarrow 0}
\frac{q\cdot N^{\ast}}{q\cdot N
q\cdot N^{\ast} + i\epsilon}, \right.
\label{eq:MLPres}
\end{eqnarray}
$N_{\mu}^{\ast} \equiv (N_0, -\bf{N})$ being the dual vector of
$N_{\mu}$.

\begin{figure}[p]
\centerline{\epsfbox{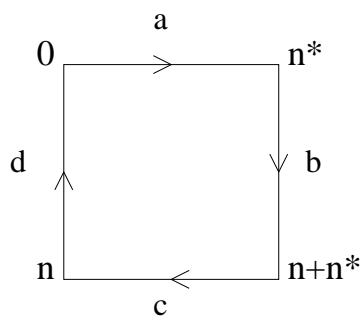}}
\vspace{1cm}
\caption{Rectangular Wilson loop with light-like segments.}
\label{fg:Loop}
\end{figure}

The purpose of this article is to test the unified-gauge formalism
in Yang-Mills theory
by evaluating the one-loop expectation value of the Wilson
loop [6,7,8,9,11,12] for the rectangular path shown in
Figure~\ref{fg:Loop}.
The path lies in Minkowski space and is charaterized in terms of
the two light-cone vectors, $n_{\mu}\equiv (n_0,\bf{n})$ and
$n_{\mu}^{\ast} \equiv (n_0,-\bf{n})$: $n^2 = (n^{\ast})^2 = 0$.
The four sides of the oriented path
from a to d are parameterized thus:
\begin{eqnarray}
x_{\mu}^a & = & n_{\mu}^{\ast} t,  t\in [0,1), \nonumber \\
x_{\mu}^b & = & n_{\mu}^{\ast} + n_{\mu}s, s\in [0,1), \nonumber \\
x_{\mu}^c & = & n_{\mu} + n_{\mu}^{\ast} u, s\in [1,0), \nonumber
\\
x_{\mu}^d & = & n_{\mu}v, v\in [1,0).
\end{eqnarray}
Notice the novel approach of using $distinct$ sets of
vectors
for the
path (5), $\{n_{\mu}, n_{\mu}^{\ast}\}$, and for the gauge-fixing
condition (2), namely $\{N_{\mu}, N_{\mu}^{\ast}\}$.

\begin{figure}[p]
\centerline{\epsfbox{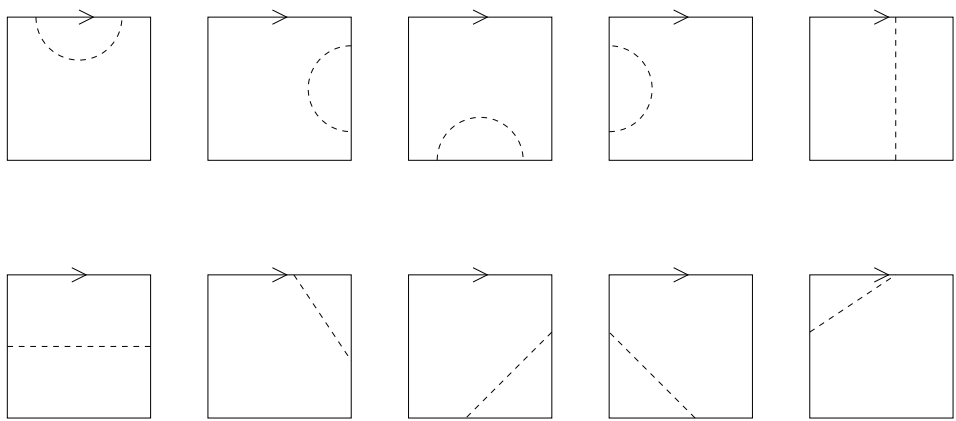}}
\vspace{1cm}
\caption{The ten first-order diagrams for the Wilson loop depicted
in Figure 1.}
\label{fg:Allten}
\end{figure}

Figure~\ref{fg:Allten} shows the ten diagrams contributing to the
first-order expectation value of the Wilson loop, $W^{(1)}$.
These diagrams lead to the following expression:
\begin{eqnarray}
W^{(1)} & = & (ig)^2 C_F \mu^{4-D} \int \frac{d^Dq}{(2\pi)^D}
G^{\mu\nu}
(q)\int_0^1 dt \int_0^1 dt^{\prime} \left[ n_{\mu}^{\ast}
n_{\nu}^{\ast}( e^{iq\cdot n^{\ast}(t-t^{\prime})}
\right. \nonumber \\
& &  - e^{-iq\cdot n^{\ast}(t-t^{\prime})+iq\cdot n})
+n_{\mu}n_{\nu} ( e^{iq\cdot n(t-t^{\prime})}-e^{-iq\cdot
n(t-t^{\prime})
-iq\cdot n^{\ast}}) \nonumber \\
& & +n_{\mu}n_{\nu}^{\ast}(e^{-iq\cdot n^{\ast}t+iq\cdot n
t^{\prime}
+iq\cdot n^\ast}- e^{-iq\cdot n t + iq\cdot n^{\ast}t^{\prime}
+iq\cdot (n-n^{\ast})} \nonumber \\
& & \left. + e^{-iq\cdot n^{\ast}t+iq\cdot n t^{\prime}-iq\cdot n}
- e^{iq\cdot n^{\ast}t-iq\cdot n t^{\prime}})\right].
\label{eq:Collect}
\end{eqnarray}

We now have to decide whether to perform $first$ the
momentum
integration and $then$ the path integrations, or whether to
begin by
integrating first over t and $t^{\prime}$.  Of course, the
traditional and generally more convenient approach has been to
start with the $d^4q$ integration (see Section 3).  But, as we
shall demonstrate in Section 2, it is also technically feasible to
begin
with the $t$, $t^{\prime}$ integrations.  As expected both
approaches
yield identical results.

\section{Performing the Path Integrations First}

Integration over the path variables t and $t^{\prime}$ in
Eq.~(\ref{eq:Collect}) yields the following intermediate result for
$W^{(1)}$:
\begin{eqnarray}
W_{path}^{(1)} & = & (ig)^2 C_F \mu^{4-D} \int
\frac{d^Dq}{(2\pi)^D}
G^{\mu\nu}(q)
\left\{ \frac{n_{\mu}^{\ast}n_{\nu}^{\ast}}{(q\cdot n^{\ast})^2}
(2 -e^{iq\cdot n^{\ast}}-e^{-iq\cdot n^{\ast}})\right. \nonumber \\
& & \times (1-e^{-iq\cdot n}) + \frac{n_{\mu}n_{\nu}}{(q\cdot n)^2}
(2-e^{iq\cdot n}-e^{-iq\cdot n})(1-e^{-iq\cdot n^{\ast}})
\nonumber \\
& & - \frac{n_{\mu}n_{\nu}^{\ast}}{q\cdot n q\cdot n^{\ast}}
[(e^{iq\cdot n} -1)(e^{iq\cdot n^{\ast}} -1)
+(e^{iq\cdot n} -1)(e^{-iq\cdot n^{\ast}} -1) \nonumber \\
& & \left. +(e^{-iq\cdot n} -1)(e^{-iq\cdot n^{\ast}} -1) +
(e^{-iq\cdot n} -1)(e^{iq\cdot n^{\ast}} -1)]\right\}.
\label{eq:pathf}
\end{eqnarray}
Notice the initial presence of the three denonimators, namely
$(q\cdot n^{\ast})^2$,
$(q\cdot n)^2$, and $(q\cdot n q\cdot n^{\ast})$.  Surprisingly,
all
three denominators disappear upon contraction of the
Lorentz indices:

\begin{eqnarray}
\frac{n_{\mu} n_{\nu}^{\ast} G^{\mu\nu}(q)}{q\cdot n q\cdot
n^{\ast}}
& = & \frac{-i}{q^2 + i\epsilon}\left[ \frac{n\cdot n^{\ast}}
{q\cdot n q\cdot n^{\ast}} - \frac{N\cdot n^{\ast}}
{q\cdot N q\cdot n^{\ast}} - \frac{N\cdot n} {q\cdot N q\cdot n}
+\frac{N^2}{(q\cdot N)^2}\right],
\nonumber \\
\frac{n_{\mu}^{\ast}n_{\nu}^{\ast} G^{\mu\nu}(q)}{(q\cdot
n^{\ast})^2}
& = & \frac{-i}{q^2 + i\epsilon}\left[ \frac{-2N\cdot n^{\ast}}
{q\cdot N q\cdot n^{\ast}} + \frac{N^2}{(q\cdot N)^2}\right],
\nonumber \\
\frac{n_{\mu} n_{\nu} G^{\mu\nu}(q)}{(q\cdot n)^2}
& = & \frac{-i}{q^2 + i\epsilon}\left[ \frac{-2N\cdot n}
{q\cdot N q\cdot n} + \frac{N^2}{(q\cdot N)^2}\right].
\label{eq:lcon}
\end{eqnarray}
When Eqs.~(\ref{eq:lcon}) are substituted into
Eq.~(\ref{eq:pathf}),
we obtain:
\begin{eqnarray}
W_{path}^{(1)} & = & (ig)^2 C_F \mu^{4-D} 2 n\cdot n^{\ast} \int
\frac{d^Dq}{(2\pi)^D}\left( \frac{-i}{q^2+i\epsilon}\right)
\frac{1}{q\cdot n q\cdot n^{\ast}} \nonumber \\
& & \times [-2 + 2 e^{iq\cdot n} + 2 e^{iq\cdot n^{\ast}} -
e^{iq\cdot (n+n^{\ast})}
- e^{iq\cdot (n-n^{\ast})}].
\label{eq:pathd}
\end{eqnarray}
The remarkable fact about Eq.~(\ref{eq:pathd}) is that there
is
$no \ dependence$ on the gauge-fixing vector $N_{\mu}$.
Our approach of using distinct vectors for the path of the Wilson
loop
($n_{\mu}$) and for the gauge-constraint ($N_{\mu}$) has allowed us
to exhibit
unambiguously the gauge invariance of the Wilson loop.

The potential pole from the $q\cdot n^{\ast}$-term in the
denominator of Eq.~(\ref{eq:pathd}) is harmless, since it is
cancelled by the numerator in the limit as
$q\cdot n^{\ast}\rightarrow 0$.
There is, however a singularity from the $q\cdot n$ pole, which may
be treated by the
prescription given in Eq.~(\ref{eq:MLPres}).

In order to perform the momentum integration in
Eq.~(\ref{eq:pathd}),
we first
parameterize the denominators as follows:
\begin{eqnarray}
\frac{1}{q^2 + i\epsilon} & = & -i\int_0^{\infty} d\alpha
e^{i\alpha(q^2+i\epsilon)},\, \epsilon > 0.
\nonumber \\
\frac{1}{q\cdot n q\cdot n^{\ast} + i\epsilon} & = & -i
\int_0^{\infty} d\beta e^{i\beta(q\cdot n q\cdot
n^{\ast}+i\epsilon)}.
\label{eq:param}
\end{eqnarray}
Substitution of the above parameterizations into Eq.~(\ref{eq:pathd})
yields the
following expression for $W^{(1)}$:
\begin{eqnarray}
W_{path}^{(1)} & = & (ig)^2 C_F \mu^{4-D} 2 n\cdot
n^{\ast}\frac{i}{(2\pi)^D}
\int_0^{\infty} d\alpha \int_0^{\infty}d\beta
e^{-(\alpha+\beta)\epsilon}
\int d^Dq e^{iq_0^2(\alpha + \beta n_0^2)}
\nonumber \\
& & \times e^{-i\alpha {\bf q}^2 -
i\beta({\bf q} \cdot {\bf n})^2} [-2+2e^{iq\cdot n} +
2e^{iq\cdot n^{\ast}} -e^{2iq_0n_0} - e^{-2i{\bf q} \cdot {\bf
n}}].
\end{eqnarray}
The momentum integration then gives us
\begin{eqnarray}
W_{path}^{(1)} & = & (ig)^2 C_F \mu^{4-D} 2 n\cdot
n^{\ast}\frac{i\pi^{D/2}}
{(2\pi)^D}\int_0^{\infty} d\alpha (i\alpha)^{1-D/2}
\int_0^{\infty}d\beta \frac{e^{-(\alpha + \beta)\epsilon}}
{\alpha+\beta n_0^2}
\nonumber \\
& & \times \left[ 2 - \exp\left(
\frac{-i n_0^2}{\alpha + \beta n_0^2}\right) - \exp\left(
\frac{i {\bf n}^2}{\alpha + \beta n_0^2}\right)\right].
\end{eqnarray}
Letting $\beta^{\prime} = \beta n_0^2$, and making the
substitution
\begin{equation}
\alpha = \lambda (1-\xi),\,\,\, \beta^{\prime} = \lambda\xi,
\end{equation}
we find that
\begin{eqnarray}
W_{path}^{(1)} & = & (ig)^2 C_F \mu^{4-D}\frac{4\pi^{D/2}}
{(2\pi)^D}i^{2-D/2}\int_0^1 d\xi (1-\xi)^{1-D/2}
\nonumber \\
& & \times \int_0^{\infty} d\lambda \frac{
e^{-\lambda(1-\xi+\xi/n_0^2)\epsilon}}
{\lambda^{D/2-1}}\left[ 2-\exp\left(\frac{-in_0^2}{\lambda}
\right) -\exp\left(\frac{i{\bf n}^2}{\lambda}\right)\right].
\label{eq:laxi}
\end{eqnarray}
Since $(1-\xi+\xi/n_0^2)> 0$, we may set $(1-\xi+\xi/n_0^2)
\epsilon = \epsilon^{\prime}$ to get for the $\xi$-integration,
\begin{equation}
\int_0^1 d\xi (1-\xi)^{1-D/2} = \frac{\Gamma (2-D/2)}{\Gamma
(3-D/2)}
= \frac{2}{4-D} + 0(4-D).
\label{eq:xiint}
\end{equation}
Hence one of the two expected poles as $D\rightarrow 4$ is provided
by
the $\xi$-integration:
\begin{eqnarray}
W_{path}^{(1)} & = & (ig)^2 C_F^{\mu^{4-D}}
\frac{8}{(4\pi)^{D/2}}\frac{i^{2-D/2}}{4-D}\int_0^{\infty}
\frac{d\lambda}{\lambda^{D/2-1}}
\nonumber \\
& & \times\left[ 2-\exp\left(\frac{-in_0^2}{\lambda}
\right) -\exp\left(\frac{i{\bf n}^2}{\lambda}\right)\right],
\nonumber \\
W_{path}^{(1)} & = & \frac{-g^2 C_F
\mu^{4-D}i^{2-D/2}}{(2\pi)^{D/2}}
\frac{4\Gamma(\frac{D}{2}-1}{(4-D)^2}
\left[ (\mu_0^2 + i\eta^{\prime})^{2-D/2} \right.
\nonumber \\
& & \left. + (-\mu_0^2 + i\eta^{\prime})^{2-D/2}\right],
\eta^{\prime} > 0.
\label{eq:paron}
\end{eqnarray}

\section{Performing the Momentum Integration First}

The first step is to apply prescription
(4) to
the gauge-field propagator in Eq.~(\ref{eq:Nprop}), setting $\alpha
= 0$:
\begin{equation}
G_{\mu\nu}(q) = \frac{-i}{q^2+i\epsilon}\left[ g_{\mu\nu}
-\frac{q\cdot N^{\ast}(q_{\mu}N_{\nu}+q_{\nu}N_{\mu})}
{q\cdot N q\cdot N^{\ast}+i\epsilon} + \frac{N^2(q\cdot N^{\ast})^2
q_{\mu}q_{\nu}}{(q\cdot N q\cdot N^{\ast}+i\epsilon)^2}\right].
\label{eq:NpropM}
\end{equation}
Substitution of Eq.~(\ref{eq:NpropM}) into Eq.~(\ref{eq:Collect}),
followed
by
a suitable re-arrangement of terms, yields the expression
\begin{equation}
W_{mom}^{(1)} = (ig)^2 C_F \frac{\mu^{4-D}}{(2\pi)^D} \sum_{i=1}^5
I_i,
\label{eq:Idef}
\end{equation}
where
\begin{eqnarray}
I_1 &=& 4iN_0^2n_0 \int_0^1 dt \int_0^1 dt^{\prime} \frac{\partial}
{\partial t} \int d^Dq \frac{-iq_0}{(q\cdot N q\cdot N^{\ast}
+ i\epsilon)(q^2+i\epsilon)}\nonumber \\
& & \times (e^{iq\cdot n(t-t^{\prime})}-e^{iq\cdot n(t-t^{\prime})
+iq\cdot n^{\ast}} + e^{iq\cdot n^{\ast}t+iq\cdot t^{\prime}}
-e^{iq\cdot n^{\ast} t -iq\cdot n t^{\prime}}),
\\
I_2 &=& -4i{\bf n}\cdot {\bf N}\int_0^1 dt \int_0^1 dt^{\prime}
\frac{\partial}{\partial t} \int d^Dq \frac{-i{\bf q}\cdot {\bf
N}}
{(q\cdot N q\cdot N^{\ast} + i\epsilon)(q^2+i\epsilon)}\nonumber \\
& & \times (e^{iq\cdot n(t-t^{\prime})}-e^{iq\cdot n(t-t^{\prime})
+iq\cdot n^{\ast}} + e^{iq\cdot n^{\ast}t+iq\cdot t^{\prime}}
-e^{iq\cdot n^{\ast} t -iq\cdot n t^{\prime}}),
\label{eq:Isub2} \\
I_3 &=& 2N^2 \int_0^1 dt \int_0^1 dt^{\prime} \frac{\partial^2}
{\partial t\,\partial t^{\prime}} \int d^Dq \frac{-iq_0^2 N_0^2}
{(q\cdot N q\cdot N^{\ast} + i\epsilon)(q^2+i\epsilon)}\nonumber \\
& & \times (e^{iq\cdot n(t-t^{\prime})}-e^{iq\cdot n(t-t^{\prime})
+iq\cdot n^{\ast}} - e^{iq\cdot n^{\ast}t+iq\cdot n t^{\prime}}
-e^{iq\cdot n^{\ast} t -iq\cdot n t^{\prime}}),
\\
I_4 &=& 2N^2 \int_0^1 dt \int_0^1 dt^{\prime} \frac{\partial^2}
{\partial t\,\partial t^{\prime}} \int d^Dq \frac{-i({\bf q}\cdot
{\bf N})^2} {(q\cdot N q\cdot N^{\ast} + i\epsilon)
(q^2+i\epsilon)}\nonumber \\
& & \times (e^{iq\cdot n(t-t^{\prime})}-e^{iq\cdot n(t-t^{\prime})
+iq\cdot n^{\ast}} - e^{iq\cdot n^{\ast}t+iq\cdot n t^{\prime}}
-e^{iq\cdot n^{\ast} t -iq\cdot n t^{\prime}}),
\\
I_5 &=& 2n\cdot n^{\ast} \int_0^1 dt\int_0^1 dt^{\prime} \int d^Dq
(e^{iq\cdot n^{\ast}t+iq\cdot n t^{\prime}}-e^{iq\cdot n^{\ast}t
-iq\cdot n t^{\prime}})\frac{-i}{q^2+i\epsilon}.
\label{eq:Isub5}
\end{eqnarray}
The contributions $I_1,\ldots,I_4$ vanish. Let us
demonstrate the vanishing of $I_2$. When the $d^Dq$ integration is
performed in Eq.~(\ref{eq:Isub2}), we obtain
\begin{eqnarray}
I_2 & = & -4i\int_0^{\infty}dt \int_0^{\infty}dt^{\prime}({\bf
n}\cdot
{\bf N})^2 \int_0^{\infty}d\alpha \int_0^{\infty} d\beta
\frac{\pi^{D/2}(i\alpha)^{1-D/2}e^{-(\alpha+\beta)\epsilon}}
{(\alpha+\beta {\vec N}^2)^{3/2}(\alpha + \beta {N_0^2})^{1/2}}
\nonumber \\
& & \times \frac{1}{2} \frac{\partial}{\partial t}\left[
-i(t-t^{\prime})e^{a(t-t^{\prime})^2+b(t-t^{\prime})^2}
+i(t-t^{\prime}) e^{a(t+t^{\prime})^2+b(t-t^{\prime})^2}
\right. \nonumber \\
& & \left. -i(1-t+t^{\prime})
e^{a(1+t-t^{\prime})^2+b(1-t+t^{\prime})^2}
-i(t+t^{\prime}) e^{a(t-t^{\prime})^2+b(t+t^{\prime})^2}\right],
\nonumber \\
\end{eqnarray}
where
\begin{equation}
a=-\frac{in_0^2}{4(\alpha+\beta N_0^2)},
\,\,\, b= \frac{i{\bf n}^2}{4\alpha}-\frac{i\beta({\bf n}\cdot
{\bf N})^2}{4\alpha(\alpha+\beta {\bf N}^2)}.
\end{equation}
Setting $\beta^{\prime} = \beta n_0^2$, and making the
substitution
\begin{equation}
\alpha = \lambda (1-\xi),\,\,\, \beta^{\prime} =\lambda \xi ,
\end{equation}
we see that
\begin{eqnarray}
I_2 & = & 2i\frac{({\bf n}\cdot {\bf
N})^2}{N_0^2}\pi^{D/2}i^{1-D/2}
\int_0^1 d\xi \frac{(1-\xi)^{1-D/2}}{(1-\xi+\xi {\bf
N}^2/N_0^2)^{3/2}}
\int_0^1 dt\int_0^1 dt^{\prime}
\nonumber \\
& & \times \int_0^{\inf} d\lambda \frac{e^{-\lambda(1-\xi
+\xi/N_0^2)
\epsilon}} {\lambda^{D/2}} \frac{\partial}{\partial t}\left[
i(t-t^{\prime}) e^{iA(t-t^{\prime})^2/\lambda
+iB(t-t^{\prime})^2/\lambda}
\right. \nonumber \\
& & +i(1-t+t^{\prime}) e^{iA(1+t-t^{\prime})^2/\lambda
+iB(1-t+t^{\prime})^2/
\lambda} - i(t-t^{\prime}) e^{iA(t+t^{\prime})^2/\lambda+
iB(t-t^{\prime})^2/\lambda} \nonumber \\
& & \left.
+i(t+t^{\prime}) e^{a(t-t^{\prime})^2/\lambda
+iB(t+t^{\prime})^2/\lambda}\right];
\end{eqnarray}
here,
\begin{equation}
A = -\frac{n_0^2}{4}, \,\,\, B= \frac{{\bf n}^2}{4(1-\xi)}
-\frac{\xi({\bf n}\cdot {\bf N})^2}{4N_0^2(1-\xi)(1-\xi+\xi
{\bf N}^2/N_0^2)}.
\end{equation}
The $\lambda$ integration yields
\begin{eqnarray}
I_2 &=& \frac{i({\bf n}\cdot {\bf N})^2}{2N_0^2} 2^D\pi^{D/2}
\int_0^1 d\xi \frac{(1-\xi)^{1-D/2}}{(1-\xi+\xi{\vec
N}^2/N_0^2)^{3/2}}
\int_0^1 dt \int_0^1 dt^{\prime} \nonumber \\
& & \times \frac{\partial}{\partial t} \left\{
\frac{i(t-t^{\prime})}{[-(A(t-t^{\prime})^2
+B(t-t^{\prime})^2)^2 + i\epsilon]^{D/2}}
\right.\nonumber \\
& & -\frac{i(t-t^{\prime}-1)}{[-(A(t-t^{\prime}+1)^2
+B(t-t^{\prime}-1)^2)^2 + i\epsilon]^{D/2}}
\nonumber \\
& & -\frac{i(t-t^{\prime})}{[-(A(t+t^{\prime})^2
+B(t-t^{\prime})^2)^2 + i\epsilon]^{D/2}}\nonumber \\
& & \left. +\frac{i(t+t^{\prime})}{[-(A(t-t^{\prime})^2
+B(t+t^{\prime})^2)^2 + i\epsilon]^{D/2}}
\right\}.
\label{eq:Murr}
\end{eqnarray}
The reader may convince himself that the $t$-integration gives the
result $I_2=0$.
Before continuing with $I_5$, we notice the curious fact that the
four integrals
$I_1,\ldots,I_4$ in Eqs. (19) - (22) all $depend$ on the
gauge-fixing vectors
$N_{\mu}, N_{\mu}^{\ast}$, while $I_5$ in Eq.~(\ref{eq:Isub5}) is
$N_{\mu}$-
independent.  Since $I_1,\ldots,I_4$ are zero, however, the
expression for the Wilson
loop $W_{mom}^{(1)}$ in Eq.~(\ref{eq:Idef}) is indeed gauge-
$independent$.
To evaluate the only non-zero contribution in Eq.~(\ref{eq:Isub5}),
we proceed by first
noting the formula [13]:
\begin{equation}
\int \frac{d^D q \, e^{ip.m}}{p^2 +i\epsilon} = \pi^{D/2}
\Gamma(\frac{D}{2}-1)
(4/m^2)^{\frac{D}{2}-1}.
\label{eq:new1}
\end{equation}
Accordingly, the two momentum integrals in Eq.~(\ref{eq:Isub5})
give
\begin{eqnarray}
\int \frac{d^Dq \, e^{i(q\cdot n^{\ast}t + q\cdot
nt^{\prime})}}{q^2 + i\epsilon}
& = & \pi^{D/2} \Gamma(\frac{D}{2}-1) (tt^{\prime} n_0^2 +
i\eta)^{1-
\frac{D}{2}}, \epsilon > 0, \eta > 0; \nonumber \\
\int \frac{d^D q \, e^{i(q\cdot n^{\ast}t - q\cdot
nt^{\prime})}}{q^2 + i\epsilon}
& = & \pi^{D/2} \Gamma(\frac{D}{2}-1) (-tt^{\prime} n_0^2 +
i\eta)^{1-
\frac{D}{2}}, \epsilon > 0, \eta > 0,
\label{eq:new2}
\end{eqnarray}
so that
\begin{eqnarray}
I_5 & = & -2i n\cdot n^{\ast}(n_0^2)^{1-\frac{D}{2}}
\pi^{\frac{D}{2}}
\Gamma(\frac{D}{2}-1)
\nonumber \\
& & \int_0^1 dt \int_0^1 dt^{\prime} \lbrace (tt^{\prime} +
i\eta^{\prime})^{1-
\frac{D}{2}} - (-tt^{\prime} +
i\eta^{\prime})^{1-\frac{D}{2}}\rbrace,
\label{eq:new3}
\end{eqnarray}
where $n\cdot n^{\ast} = 2n_0^2$ and $\eta^{\prime} =
\eta/\eta_0^2$.  The integration
over $t$ and $t^{\prime}$ is easy and leads, in the limit as
$D\rightarrow 4$, to
\begin{equation}
I_5 = \frac{-16 \pi^{\frac{D}{2}} \Gamma (\frac{D}{2}-1)}{(4-D)^2}
\left[(n_0^2 +
i\eta^{\prime})^{2-\frac{D}{2}} + (-n_0^2 +
i\eta^{\prime})^{2-\frac{D}{2}} \right],
\eta^{\prime} > 0.
\label{eq:new4}
\end{equation}
Substituting the result~(\ref{eq:new4}) into the expression for the
Wilson loop
$W_{mom}^{(1)}$, Eq.~(\ref{eq:Idef}), we finally obtain
\begin{eqnarray}
W_{mom}^{(1)} & = & (ig)^2 C_F \frac{\mu^{4-D}}{(2\pi)^D} I_5,
\nonumber \\
W_{mom}^{(1)} & = & \frac{+4ig^2 C_F \mu^{4-D} \Gamma(\frac{D}{2}-
1)}{(2\pi)^{\frac{D}{2}} (4-D)^2} \nonumber \\
& & \left[ (n_0^2 +
i\eta^{\prime})^{2-\frac{D}{2}} + (-
n_0^2 + i\eta^{\prime})^{2-\frac{D}{2}} \right], \eta^{\prime} > 0.
\label{eq:new5}
\end{eqnarray}
This answer agrees with Eq. (3.1) in ref. [3].

Comparing the result (\ref{eq:new5}) with $W_{path}^{(1)}$ in
Eq.~(\ref{eq:paron}),
we see that the two distinct integration sequences give identical
results (the inessential
factor $i^{2-\frac{D}{2}}$ in Eq.~(\ref{eq:paron}) reduces to unity
as $D\rightarrow
4$).

\section{Discussion}

In this paper we have demonstrated the gauge independence of the
Wilson loop to one-
loop order for a general class of axial-type gauges.  Our final
results are listed in
Eqs.~(\ref{eq:paron}) and (\ref{eq:new5}).  Working in the
unified-gauge formalism,
characterized by the fixed four-vector $N_{\mu}$, we were able to
convince ourselves
that all integrations were ambiguity-free, regardless of the nature
of $N_{\mu}$, and
regardless of the order of integration.

To assist us in our analysis we decided to use $distinct$
sets
of vectors for the paths,
$\{n_{\mu}, n_{\mu}^{\ast} \}$, and for the gauge-fixing
constraint, $\{ N_{\mu},
N_{\mu}^{\ast} \}$.  With the help of this technical "fine-tuning",
we showed that the
correct result (Eq.~(\ref{eq:paron}), or Eq.~(\ref{eq:new5})) could
be obtained, either
by integrating $first$ over the path variables $t$ and
$t^{\prime}$ and then
over the momentum variables $d^4q$ (cf. $W_{path}^{(1)}$), or by
$first$
integrating over the momenta (cf. $W_{mom}^{(1)}$).  Judging from
the specifics of
each calculation, it would appear that the procedure leading to
$W_{path}^{(1)}$ is
shorter and, perhaps, wrought with fewer difficulties, than the
approach for
$W_{mom}^{(1)}$.

\section{Acknowledgments}

The authors should like to thank Stefano Fachin for numerous
discussions and for his continued interest in this project.
One of us (B.J.H.) would like to thank the Natural Sciences and
Engineering Research Council of Canada for an NSERC Postdoctoral
Fellowship. This work was supported in part by the Natural Sciences
and Engineering Research Council of Canada under Grant No. A 8063.


\begin{thebibliography}{99}

\bibitem{cc:pa0g} S. Caracciolo, G. Curci and P. Menotti, Phys.
Lett. B113 (1982) 311.
\bibitem{at:pbML} A. Andr\v{a}si and J.C. Taylor, Nucl. Phys. B375
(1992) 341; Nucl. Phys. B414 (1994) E856.
\bibitem{bk:giad} A. Bassetto, I.A. Korchemskaya, G.P. Korchemsky
and G. Nardelli, Nucl. Phys. B408 (1993) 62.
\bibitem{ma:2} S. Mandelstam, Nucl. Phys. B213 (1983) 149.
\bibitem{le:3} G. Leibbrandt, Phy. Rev. D29 (1984) 1699.
\bibitem{le:gptp} G. Leibbrandt, Nucl. Phys. B310 (1988) 405.
\bibitem{le:1} G. Leibbrandt, $Nonconvariant \ Gauges$
(World Scientific, Singapore, 1994).
\bibitem{l:ugft} G. Leibbrandt, Nucl. Phys. B337 (1990) 87.
\bibitem{hu:1} H\H{u}ffel, P.V. Landshoff and J.C. Taylor, Phys. Lett. B217
(1989) 147.
\bibitem{kk:1} I.A. Korchemskaya and G.P. Korchemsky, Phys. Lett. B287 (1992)
169.
\bibitem{la:1} P.V. Landshoff, Phys. Lett. 169B (1986) 69.
\bibitem{b:4} A. Bassetto, G. Nardelli and R. Soldati,
{\twlit Yang-Mills theories in algebraic non-covariant gauges} (World
Scientific, Singapore, 1991).
\bibitem{pa:1} P. Pascual and R. Tarrach,
{\twlit QCD: Renormalization for the Practitioner},
Lecture Notes in Physics,
Vol. 194, Springer-Verlag (Berlin, Heidelberg, 1984).
\end{thebibliography}
\end{document}